# Vertical Nb Josephson junctions fabricated by direct metal deposition on both surfaces of freestanding graphene layers

*Yoonkang Kim[1,2,†], Seongbeom Kim[1,2,†], Jeonglyul Kim[3,4], Kikyung Jung[1], Sejin An[3,4], Jieun Lee[1], Hyobin Yoo[3,4], Joon Young Park[2,5,\*], Gyu-Chul Yi[1,\*]*

[1] Department of Physics and Astronomy, Institute of Applied Physics, Seoul National University, 1, Gwanak-ro, Gwanak-gu, Seoul, 08826, Republic of Korea

[2] Center for 2D Quantum Heterostructures, Institute for Basic Science (IBS), Sungkyunkwan University (SKKU), Suwon 16419, Republic of Korea

[3] Department of Materials Science and Engineering, Seoul National University, 1, Gwanak-ro, Gwanak-gu, Seoul, 08826, Republic of Korea

[4] Research Institute of Advanced Materials, Seoul National University, 1, Gwanak-ro, Gwanak-gu, Seoul, 08826, Republic of Korea

[5] Department of Physics, Sungkyunkwan University (SKKU), Suwon 16419, Republic of Korea

[†] These authors contributed equally to this work.
\* Author to whom correspondence should be addressed.
Contact E-mail: park.jy@skku.edu; gcyi@snu.ac.kr

**Abstract**

Vertical integration of superconducting electronics requires fabrication strategies that preserve pristine interfaces while accommodating oxidation-sensitive elemental superconductors. However, existing van der Waals–based vertical Josephson junctions largely rely on transfer-based assembly schemes that are incompatible with elemental materials such as niobium (Nb). Here, we introduce a freestanding van der Waals membrane architecture that enables deposition-based fabrication of vertical Josephson junctions through double-sided processing of a single suspended two-dimensional layer. Using multilayer graphene suspended across lithographically defined through-holes in a SiN$_x$ membrane, we realize vertical Nb/multilayer graphene/Nb Josephson junctions without ambient exposure of buried interfaces. The resulting devices exhibit clear Josephson coupling, including reproducible supercurrents and a temperature dependence of the critical current consistent with short-junction behaviour. Well-defined magnetic interference patterns governed by the membrane-defined aperture geometry, together with sub-gap features that track a Bardeen–Cooper–Schrieffer (BCS)-like superconducting gap, further confirm the junction quality. This platform establishes a scalable route to vertical superconducting devices based on oxidation-sensitive elemental superconductors and van der Waals materials.

## Introduction

Freestanding membrane–based device architectures offer a distinct route to vertical electronic integration by enabling junction formation in a geometry that is defined architecturally rather than by post-patterning on a supporting substrate.[1-4] By suspending atomically thin van der Waals (vdW) materials across lithographically defined through-hole apertures, these architectures provide independent, sequential access to the active region from both sides along the out-of-plane direction.[1] This allows both junction geometry and interface formation to be set by the membrane aperture itself, reducing reliance on alignment-sensitive overlap processes. A defining feature is that the suspended vdW layer functions not only as an active electronic element but also as a conformal, atomically thin capping layer during sequential processing from opposite sides.[1,3,4] This intrinsic capping mechanism suppresses oxidation and chemical degradation of buried interfaces without requiring in-situ encapsulation or glove-box–based handling.[1] The generality of this concept has been demonstrated previously in double-sided van der Waals epitaxial growth of three-dimensional topological insulators across suspended $SiN_x$ membranes, establishing that symmetric, substrate-free processing can yield high-quality quantum electronic heterostructures with pristine interfaces.[1] Extending this architectural principle from materials growth to general electronic and superconducting device fabrication therefore offers a versatile platform in which device geometry, interface quality, and integration density are deterministically defined at the earliest stage of fabrication.

Leveraging the advantages of freestanding membrane–based device architectures described above, we address fundamental limitations in existing superconducting quantum technologies.

Superconducting quantum circuits are built upon Josephson junctions, which have been predominantly realized using Al/AlO$_x$/Al tunnel junctions.[5-7] While this materials system has enabled reproducible large-scale integration, the low superconducting transition temperature of Al restricts device operation to extremely low temperatures.[5-9] Moreover, the amorphous nature of oxide barriers imposes intrinsic constraints on materials flexibility, while the presence of two-level defects in amorphous barriers introduces decoherence pathways that are detrimental to quantum information processors.[10-12] Extending the operating temperature range and improving device performance therefore require the use of superconductors with higher transition temperatures, such as niobium (Nb). However, integrating elemental superconductors such as Nb into Josephson junctions is challenging due to the difficulty of forming a well-defined non-superconducting barrier layer. In particular, conventional amorphous oxide barriers are incompatible with Nb, as uncontrolled oxidation of the Nb electrode leads to significant interfacial disorder. To overcome this barrier-layer bottleneck, van der Waals (vdW) materials have been explored as alternative non-superconducting layers in vertical Josephson junctions, owing to their atomically flat, dangling-bond-free interfaces[13,14] and their ability to support proximity-coupled transport.[15-19] Despite these advantages, most previously reported vdW-based vertical Josephson junctions rely on transfer-based stacking of vdW layers and superconducting electrodes, often requiring specialized handling environments.[16-19] Such approaches limit scalability, complicate integration with standard microfabrication workflows, and remain incompatible with oxidation-sensitive elemental superconductors such as Nb.[20-22] This incompatibility directly limits the realization of Nb-based vertical Josephson junctions within fabrication schemes compatible with superconducting integrated circuits.

In this article, we demonstrate that a freestanding vdW membrane architecture enables deposition-based fabrication of vertical Josephson junctions incorporating oxidation-sensitive elemental superconductors. Using multilayer graphene suspended across lithographically defined through-holes in an in-house–fabricated freestanding SiN$_x$ membrane, we realize vertical Nb/multilayer graphene/Nb Josephson junctions following the double-sided processing sequence. In this architecture, the suspended multilayer graphene simultaneously functions as a proximity-coupled weak link and as a conformal capping layer that protects the first-deposited electrode during subsequent top-side processing. This bidirectional fabrication strategy enables superconductor/vdW interfaces to be formed without ambient exposure, allowing reliable Josephson coupling using sputtered Nb electrodes. The resulting devices exhibit clear Josephson supercurrents, temperature-dependent critical currents consistent with short-junction behaviour, and magnetic interference patterns governed by the membrane-defined circular junction geometry.

## Results and Discussion

### Device architecture and junction geometry

Figure 1a illustrates the freestanding membrane architecture employed to realize vertical Josephson junctions in this work. As shown in Fig. 1a(i), lithographically defined through-holes in a freestanding $SiN_x$ membrane spatially confine the active junction area, enabling vertical device geometries that are defined at the architectural level rather than by electrode overlap or post-fabrication patterning. The suspended van der Waals (vdW) layer spanning each aperture is accessible from both sides of the membrane, allowing independent top- and bottom-side processing.

Figure 1a(ii) details the fabrication strategy used to implement this concept. Multilayer graphene is first suspended across the through-hole apertures, after which superconducting Nb electrodes are sequentially deposited on the bottom and top sides of the membrane. In this geometry, the suspended graphene simultaneously functions as a proximity-coupled weak link and as a conformal capping layer that protects the initially deposited electrode during subsequent processing. A cross-sectional transmission electron microscopy image of a completed junction is shown in Fig. 1b, confirming that the multilayer graphene spans the membrane opening and is sandwiched between Nb electrodes deposited from opposite sides.

The representative junction used for transport characterization has the following geometry. The through-hole spanned by the multilayer graphene was patterned as a circular aperture with a diameter of 1 μm. Nb electrodes with a thickness of ~47 nm were deposited on both the top and bottom sides of the graphene. In particular, the top Nb electrode has a circular geometry in the top view with a diameter of approximately 2.6 μm, which is larger than the through-hole. The bottom Nb electrode also adopts a circular geometry in the bottom view, as the Nb deposition is laterally confined by the surrounding SiN$_x$ membrane owing to its thickness of ~200 nm. As a result, the vertical junction formed in this work possesses an overall cylindrical geometry (see Extended Data Fig. 1a-e). All transport measurements were performed using the configuration schematically illustrated in Fig. 1b.

## Josephson coupling and current–voltage characteristics

The basic transport characteristics of a representative junction are summarized in Figure 2. As shown in Fig. 2a, the resistance decreases gradually below ~8 K, reflecting the superconducting transitions of the top and bottom Nb electrodes, and then drops sharply to nearly zero below ~4.3 K, marking the onset of Josephson coupling. Figure 2b displays IV curves of the same device measured at temperatures between 2 K and 8 K. The supercurrent branch disappears above ~4.3 K, consistent with the junction's transition temperature. At 2 K, the device exhibits a clear Josephson supercurrent: the current–voltage (IV) characteristics display a dissipationless branch up to a critical current $I_c \approx 110\ \mu A$, followed by a transition to a resistive branch with a normal-state resistance $R_N \approx 3.6\ \Omega$ as shown in the inset of Fig. 2b. From the measured sub-gap features (see Fig. 4b), we extract a superconducting gap $\Delta_0 = 0.88$ meV ($\Delta_0 = 2\Delta_t\Delta_b/(\Delta_t + \Delta_b)$, due to unidentical gap energies of top ($\Delta_t$) and bottom ($\Delta_b$) Nb electrodes).[15] The corresponding $eI_cR_N$ product is about 0.4 meV and gives a ratio $eI_cR_N/\Delta_0 \approx 0.46$, which lies in the moderate range compared with previously reported vertical Josephson junctions.[15,17,23,24]

The IV curves measured under opposite current sweep directions overlap almost perfectly, exhibiting negligible hysteresis as shown in the inset of Fig. 2b. Given the junction geometry, the planar aperture is defined by a circular through-hole with a diameter of 1 μm, corresponding to an effective junction area of $\pi \cdot (0.5\ \mu m)^2 \approx 7.854 \times 10^{-13} m^2$. Using a typical areal capacitance of $5\ \mu F/cm^2$, the total junction capacitance is estimated to be $C \approx 39$ fF.[14,25,26] With these parameters, the Stewart–McCumber parameter can be estimated as $\beta_c = \frac{2\pi I_c R_N^2 C}{\Phi_0}$, where $\Phi_0$ is the magnetic flux quantum. Substituting the measured values yields $\beta_c \approx 0.169$, which is well below unity.[15-17] This result is consistent with overdamped junction behaviour, likely arising from effective shunting by the normal-conducting graphene layers.[15-17] Accordingly, the device behaves as a moderately transparent SNS-type Josephson junction. For the reproducibility, Extended Data

Fig. 3a, b show that an additional junction exhibits similar resistance–temperature behaviour with nearly identical critical temperature and a comparable $eI_cR_N \approx 0.5$ meV. The typical number of graphene layers was 5-6 for both junctions, based on optical contrast.

**Temperature dependence of the critical current**

To quantify the temperature dependence of the Josephson coupling, we extract $I_c(T)$ from the differential resistance maps. Fig. 2c displays a colour map of *dV/dI* as a function of bias current and temperature. The boundary between the zero-resistance region and the normal-resistance region defines the critical current at each temperature. We determine the critical current by locating, for each temperature slice, the lowest current at which *dV/dI* exceeds towards a normal resistance for several consecutive points. The resulting $I_c(T)$ envelope (symbols) is plotted in Fig. 2d. We compare the experimental $I_c(T)$ with three standard theoretical limits for SNS Josephson junctions, all based on a BCS temperature-dependent gap $\Delta(T)$. The gap is modeled using the universal formula for the BCS gap, $\Delta(T) = \Delta_0 \tanh\left[1.74\sqrt{\frac{T_c}{T} - 1}\right]$ where $\Delta_0$ is the zero-temperature gap from the experimental value ($\Delta_0 \approx 0.88$ meV for our case) and $T_c$ is the critical temperature.

The short-ballistic limit (KO-2) is described by the Kulik–Omelyanchuk expression follows Eq.1,

$$I_c^{(KO2)}(T) \propto \frac{\pi\Delta(T)}{2e}\tanh\left[\frac{\Delta(T)}{2k_BT}\right] \quad (1)$$

which applies to a perfect transparent, short junction with ballistic transport.[27,28] The short-diffusive limit (KO-1) is obtained from the Matsubara summation as shown in Eq.2 where $\omega_n = (2n+1)\pi k_B T$,

$$I_c^{(KO1)}(T) \propto \frac{\pi k_B T}{e}\sum_{n=0}^{\infty}\frac{\Delta(T)^2}{\omega_n^2 + \Delta(T)^2} \quad (2)$$

appropriate for a diffusive weak link shorter than the coherence length.[29,30] To compare the long-diffusive SNS behaviour with our junction, we additionally fitted the data using the universal scaling form of the critical current, treating the Thouless temperature as a fitting parameter.[31] However, as shown in the figure, the long-diffusive model predicts a characteristic concave temperature dependence that deviates significantly from the experimentally observed convex $I_c(T)$ envelope. The KO-1 and KO-2 best-fit curves, calculated using the same fitting parameters—$T_c$ and a global scaling factor for the critical current—are superposed on the data in Fig. 2c, d. For both models, the best-fit value of the critical temperature is $T_c \approx 4.25$ K. The required scaling factor that relates the theoretical $I_c$ to the measured values is below 10% for both KO-1 and KO-2, which is reasonable given potential sources of deviation such as instrumental noise or Fermi-velocity mismatch at the superconductor/normal-metal interface.[15,32] Across the entire temperature range, the experimental $I_c(T)$ envelope lies between the ballistic and diffusive limits. Considering this result together with the previously obtained ratio $eI_cR_N/\Delta_0 \approx 0.46$, the junction operates in the short-junction regime with moderate transparency. This behaviour is consistent with a multilayer graphene barrier consisting of 5–6 layers, whose thickness remains smaller than the superconducting coherence length of Nb, while being comparable to the out-of-plane mean free path, placing the junction in a crossover regime that is neither purely diffusive nor fully ballistic.[33-35]

**Magnetic interference and effective junction geometry**

Josephson coupling across the weak link is further supported by the magnetic-field dependence of the critical current. Figure 3 displays *dV/dI* as a function of bias current and magnetic field applied parallel to the circular junction plane (see Extended Data Fig. 1f). The device exhibits clear Fraunhofer patterns centered around zero field, with well-defined lobes and nodes, as expected for a Josephson junction. Since our junction is defined by the circular through-hole in the $SiN_x$ membrane, the relevant junction aperture is circular (axisymmetric) rather than rectangular. We therefore compare the measured interference pattern with analytic expressions for circular and rectangular geometries.[36] For a circular aperture, the critical current is described by Eq.3, where $\kappa = \frac{2\pi B t_m a}{\Phi_0}$.

$$I_c^{(circ)}(B) = I_{c0} \left| \frac{2J_1(\kappa)}{\kappa} \right| \quad (3)$$

Here, $J_1$ is the first-order Bessel function, $a$ is the junction radius, $t_m$ is the effective thickness, which includes the London penetration depth of both top, bottom Nb electrodes and the thickness of multilayer graphene. Since the thickness of the Nb electrodes is comparable to the London penetration depth of bulk Nb, we approximate the effective magnetic thickness as $t_m \approx 2 \times \lambda_L \tanh(d_{Nb}/2\lambda_L) + d_{Gr}$, where $d_{Nb}$ = 47 nm, the multilayer graphene thickness $d_{Gr}$ is negligible, and $\lambda_L$ is the estimated London penetration depth.[37] Here, $\Phi_0$ denotes the superconducting flux quantum.

For comparison, we also consider a rectangular junction of equal nominal area. Approximating it as a square of side length $L=2a$, the Fraunhofer envelope is given by Eq.4,

$$I_c^{(rect)}(B) = I_{c0} \left| \frac{\sin \kappa'}{\kappa'} \right| \quad (4)$$

with $\kappa' = \frac{\pi B t_m L}{\Phi_0}$. Figure 3 and Extended Data Fig. 2 show these two analytic envelopes overlaid on the measured $dV/dI$ map and extracted $I_c$. The circular-aperture (Bessel) model shows good agreement with the measured Fraunhofer pattern, whereas the rectangular model does not accurately reproduce the observed node positions. Within the circular-aperture model, the nodes occur at $\kappa = \alpha_n$ where $\alpha_n$ corresponds to the $n$th zero of $J_1(\kappa)$. The experimentally observed node positions are consistent with a through-hole radius of $a \approx 0.5$ µm and an effective magnetic thickness of $t_m \approx 50$ nm, in close agreement with the value expected from a London penetration depth of $\lambda_L \approx 80$ nm ($t_m \approx 45.7$ nm).

The circular (uniform-field) fit employed here implicitly assumes symmetric electromagnetic screening by the superconducting electrodes, as well as a spatially homogeneous magnetic field across the SNS weak link. In practical devices, however, geometry-dependent screening currents and flux-focusing redistribute the local magnetic field, so that the effective field at the weak link can become non-uniform and, when the electrode dimensions are dissimilar, potentially asymmetric.[38-40] Such geometry-induced distortions are known to modify the magnetic diffraction pattern and may therefore account for the residual mismatch between the idealized fit and the measured data.[38-40] Once again, Extended Data Fig. 3c shows the Fraunhofer pattern for a second device, together with the same circular and rectangular fits. Consistently, the circular-aperture model captures the main features of the interference pattern, confirming that the membrane-defined cylindrical geometry is a robust and reproducible characteristic of our vertical Josephson-junction platform.

**Sub-gap structure and multiple Andreev reflections**

To gain further insight into the junction transparency and superconducting gap, we examine the differential conductance *dI/dV* in the sub-gap regime. Fig. 4a shows representative *dI/dV* spectra measured at selected temperatures between 2 K and 4.5 K. At the lowest temperature, pronounced sub-gap features symmetric about zero bias are observed. In particular, conductance peaks indicated by red arrows appear at voltages suggestive of multiple Andreev reflection (MAR)-related subharmonic gap structure (SGS), occurring near $eV \approx 2\Delta_0/n$. The full temperature evolution of these features is summarized in Fig. 4b, which presents a two-dimensional map of *dI/dV* as a function of bias voltage and temperature. As the temperature increases toward the junction critical temperature $T_c$, both the zero-bias conductance enhancement and the sub-gap features gradually weaken and eventually disappear, consistent with the closing of the superconducting gap.

To quantify this behaviour, we overlay the expected SGS trajectories for $n = 1$ and $n = 3$ using a BCS-like temperature dependence of the superconducting gap. In the analysis, we fix $T_c \approx 4.25$ K and adopt a zero-temperature gap $\Delta_0 \approx 0.88$ meV ($2\Delta_0/e \approx 1.76$ mV). The expected MAR-related peak positions follow $V_n(T) = 2\Delta(T)/ne$, where $\Delta(T)$ is modeled using the universal formula for BCS gap employed in the $I_c(T)$ analysis.[15,17] The calculated $V_1(T)$ and $V_3(T)$ trajectories (dashed lines in Fig. 4b) closely track the experimentally observed conductance ridges over the full temperature range.

A pronounced second-order ($n = 2$) subharmonic gap structure (SGS) is not resolved in our measurements. The absence of specific subharmonic orders has previously been reported in SNS-type Josephson junctions and attributed to several intrinsic mechanisms. In SNS junctions with intermediate transparency, strong inelastic scattering renders MAR processes incoherent, leading to a complete suppression of SGS in differential conductance.[41] Even when MAR-related features persist, disorder in the weak link can strongly broaden Andreev bound states and smear sub-gap spectral features. Theoretical studies of disordered graphene Josephson junctions have shown that intervalley scattering, impurity potentials, or pseudomagnetic barriers suppress and broaden Andreev-bound-state signatures, making individual MAR-related peaks increasingly difficult to resolve.[42]

Furthermore, theoretical analyses of Josephson junctions under strong energy relaxation demonstrate that MAR consists of coherent and incoherent contributions associated with distinct quasiparticle propagators.[43] In this regime, inelastic scattering exponentially suppresses the coherent, time-dependent propagator, while the incoherent contribution remains finite and dominates transport. Consequently, the conventional picture of coherent, energy-conserving MAR ladders breaks down, and the SGS arises from higher-order corrections to the quasiparticle distribution function associated with incoherent MAR processes. Within this, the visibility of SGS features becomes strongly order dependent, and an apparent suppression of even-order subharmonic features—including the absence of the $n = 2$ threshold—naturally emerges once coherent contributions are quenched, despite the persistence of lower-order sub-gap signatures.[43]

Finally, this observation is qualitatively consistent with theoretical analyses of asymmetric superconducting weak links, which predict an intrinsic difference in the visibility of even- and odd-order MAR features.[44] In such models, odd-order resonances can receive contributions from both electron and hole branches, whereas even-order resonances are restricted to a single carrier branch, resulting in a reduced net current contribution and weaker subharmonic features.[44] Although the present junction employs nominally identical Nb electrodes on both sides, a strict symmetry between the two superconducting interfaces is not guaranteed in practice. In particular, the bottom Nb electrode is deposited first and subsequently subjected to multiple fabrication steps before the top Nb electrode is formed, which can lead to differences in interface transparency, interfacial disorder, or effective gap profiles. Such process-induced asymmetry may therefore further diminish the visibility of even-order subharmonic features, providing an additional, secondary mechanism contributing to the absence of a resolvable $n = 2$ SGS peak in our measurements.

Taken together, the selective appearance of SGS features at $n = 1$ and $n = 3$, their symmetric voltage positions, and their temperature evolution tracking the superconducting gap are all consistent with MAR-like transport in a proximity-coupled Nb-based SNS junction. While we conservatively refrain from assigning a complete MAR order sequence, the observed sub-gap structures provide suggestive evidence for Andreev-mediated transport and are fully consistent with the extracted gap and critical temperature obtained from the $I_c(T)$ analysis.

**Structural and compositional analysis of a non-functional junction**

Finally, we examine a non-functional vertical device fabricated on the same membrane platform that showed no measurable Josephson coupling. Extended Data Fig. 4a presents a HAADF–STEM cross-section of the junction. The 8-layer graphene spacer is clearly resolved between the Nb electrodes; however, additional contrast at the upper Nb/graphene interface indicates an unintended interfacial layer that is absent at the lower interface. STEM–EDS elemental mapping reveals carbon- and oxygen-rich signals extending into the upper electrode region, i.e., beyond the intrinsic graphene stack. As shown in Extended Data Fig. 4b, an anomalous ~2-nm-thick region with enhanced carbon concentration appears immediately above the graphene layers. Above this carbon-rich region, oxygen and Nb signals overlap over a thickness of ~6 nm. The corresponding atomic-fraction line profiles corroborate these trends, confirming localized enrichment of C near the graphene surface and co-localized O and Nb at the upper interface, consistent with formation of an interfacial C–O–Nb layer.

These observations indicate that the fabrication workflow can introduce interfacial contamination that suppresses the yield of functional vertical Josephson junctions. Following ex-situ low-vacuum annealing of the transferred multilayer graphene, the bottom Nb electrode is deposited directly onto the graphene surface, yielding an atomically intact lower interface, as supported by our structural and compositional characterization. By contrast, defining the top electrode requires e-beam lithography on the graphene-covered surface, exposing the upper graphene surface to polymer residues before Nb deposition. This contamination pathway is consistent with the STEM–EDS results: carbon enrichment is observed above the graphene with no detectable Nb signal, implying incomplete physical contact between the deposited Nb and the graphene. In addition, carbon residues can act as oxygen reservoirs; oxygen incorporated at the interface may react with Nb during sputtering, promoting interfacial oxidation and potentially degrading the superconducting properties of the Nb. Taken together, this comparison highlights the importance of interface cleanliness—particularly at the top electrode—for achieving high-yield vertical Josephson junctions and enabling multi-junction architectures such as SQUIDs. These findings further motivate the incorporation of residue-free patterning strategies, such as stencil-mask approaches using pre-patterned resist layers,[45,46] into the membrane-based platform to improve device reliability.

### Architectural implications for vertical Josephson junction technologies

The Nb/multilayer graphene/Nb junctions demonstrated in this work operate in a proximity-coupled SNS regime, in contrast to conventional Al/AlO$_x$/Al Josephson junctions based on SIS tunneling, making direct one-to-one quantitative performance comparisons inappropriate due to their distinct transport mechanisms. Instead, we focus on the structural and materials-level implications of the demonstrated architecture. In oxide-barrier-based junctions, fabrication convenience comes at the cost of a fundamental materials constraint: the use of amorphous oxides hinders the incorporation of elemental superconductors with higher transition temperatures, such as niobium (Nb), owing to uncontrolled interfacial oxidation and disorder.[20-22] While van der Waals (vdW) materials have been explored as alternative barrier layers, most previously reported vdW-based vertical Josephson junctions rely on transfer-based stacking, which is incompatible with deposition-based fabrication flows and oxidation-sensitive elemental superconductors.[16-19] The freestanding membrane architecture presented here directly addresses this long-standing bottleneck by enabling sequential bottom- and top-side depositions on a suspended vdW layer, which simultaneously functions as a crystalline non-superconducting barrier and as an atomically thin capping layer. While Nb/multilayer graphene/Nb junctions are shown as a representative demonstration, the key advance is the establishment of a general, oxide-free and deposition-compatible platform for vertical Josephson junctions, naturally extensible to other vdW materials—including insulating and semiconducting layers such as hBN and transition metal dichalcogenides[18,19]—and capable of supporting both proximity-coupled SNS and tunnel-type SIS junctions within the same architectural framework.

Beyond establishing a deposition-compatible route to Nb-based vertical Josephson junctions, this architectural framework also carries important implications for junction definition and circuit-level scalability. Because electrical path between the top and bottom electrodes occurs exclusively through the suspended through-hole region, the active junction area and junction density are defined a priori by the aperture size, pitch, and layout of the membrane, independent of electrode overlap or post-deposition patterning. This aperture-defined geometry allows processing on the top and bottom surfaces of the $SiN_x$ membrane to be carried out independently: bottom electrodes may be deposited uniformly over large areas, while the number, spacing, and density of vertical junctions are determined solely by the through-hole array defined on the membrane. This separation of geometric definition from electrode patterning provides a scalable route toward dense and uniform arrays of vertical Josephson junctions and enables circuit concepts that rely on out-of-plane connectivity, including SQUID-based interferometric circuits and multilayer Josephson architectures. Furthermore, by extending lithographic patterning to the backside of the membrane, individually addressable vertical junction elements can be envisioned, providing an additional degree of freedom for circuit integration and control.[47]

Moreover, the aperture-defined confinement inherent to the freestanding membrane platform is naturally compatible with future implementations using large-area, CVD-grown van der Waals materials.[1] By defining the active junction region at the membrane level, this approach enables device dimensions that are comparable to—or smaller than—typical grain sizes in CVD-grown films, thereby offering a practical strategy to mitigate grain-boundary-induced variability while preserving scalable fabrication pathways.

## Conclusions

In summary, we have demonstrated a freestanding van der Waals membrane architecture that enables oxidation-free, deposition-based fabrication of vertical Josephson junctions using elemental superconductors. By employing suspended multilayer graphene as both a crystalline non-superconducting barrier and an atomically thin capping layer, Nb/multilayer graphene/Nb junctions are realized through true double-sided deposition without transfer-based stacking or ambient exposure of buried interfaces. The resulting devices exhibit clear Josephson coupling, including reproducible supercurrents, short-junction-regime temperature dependence of the critical current, geometry-defined magnetic interference patterns, and sub-gap features that track a BCS-like superconducting gap. Crucially, the active junction area and geometry are defined by membrane apertures rather than electrode overlap, allowing independent top- and bottom-side processing and providing a scalable route to dense arrays of vertical junctions. More broadly, we believe this work establishes a general, oxide-barrier-free junction architecture for vertical superconducting heterostructures that overcomes key limitations of conventional amorphous-barrier technologies. The freestanding membrane approach is naturally extensible to other van der Waals materials and elemental superconductors, opening new opportunities for vertically integrated superconducting devices and circuits beyond oxide-based architectures.

## Materials and methods

### Preparation of suspended multilayer graphene

In-house–fabricated freestanding $SiN_x$ membranes were fabricated following a process analogous to that used for TEM window chips.[1,48] A 200-nm-thick $SiN_x$ film was first deposited on both sides of a 200-µm-thick Si (100) wafer by low-pressure chemical vapor deposition (LPCVD). The $SiN_x$ on the back side of the wafer was then selectively patterned by conventional photolithography and etched by $CF_4$ reactive-ion etching (RIE). Subsequently, the exposed Si substrate was removed by KOH wet etching, resulting in a freestanding $SiN_x$ window with no underlying support. Circular apertures were then defined on the top-side $SiN_x$ surface by electron-beam lithography and etched using $CF_4$/Ar RIE, forming an array of through-holes typically 1–3 µm in diameter. These holes serve as apertures across which the vdW layers are suspended.

Multilayer graphene was obtained by mechanical exfoliation onto a polypropylene carbonate (PPC)-coated glass slide. A PPC/polydimethylsiloxane (PDMS) stamp was then fabricated and used to transfer the graphene layers onto an oxygen-plasma-treated $SiN_x$ membrane, such that they covered several of the predefined apertures. The PPC layer was dissolved at room temperature by sequentially immersing the sample in anisole, acetone, and isopropanol. To further reduce polymer residues and improve interface cleanliness, the membrane was subsequently annealed (typically at 300 °C) under ~$10^{-4}$ Torr, yielding a clean suspended graphene region anchored to the SiN$_x$ membrane.

### Fabrication of vertical Josephson junctions

Device fabrication on suspended multilayer graphene supported by SiN$_x$ membrane templates proceeded as follows. Nb/Au (47/20 nm) was first deposited on the bottom side of the membrane by DC magnetron sputtering. The base pressure of the sputtering chamber was ~$8 \times 10^{-8}$ Torr, and Nb and Au were sputtered in an Ar atmosphere at a pressure of 3mTorr using powers of 200 W and 30 W, respectively. An in-situ Au capping layer was deposited to prevent oxidation of the underlying Nb. Subsequently, an additional Ti/Au (10/240 nm) layer was deposited by electron-beam evaporation to ensure reliable electrical contact and to provide additional mechanical support to the membrane template. Using conventional electron-beam lithography (EBL) with a bilayer resist stack (MMA/PMMA), the top junction area was patterned and developed into a circular opening slightly larger than the pre-defined SiN$_x$ through-hole (typically ~1.6 µm larger in diameter). Nb/Au (47/30 nm) was then sputtered to form the top superconducting electrode. To electrically isolate each device into an individual Josephson junction, the exposed regions of the multilayer graphene were removed by Ar plasma etching. Finally, an additional EBL step followed by electron-beam evaporation of Ti/Au (10/140 nm) was used to define the interconnection lines and bonding pads. All fabrication steps are illustrated in Extended Data Fig. 1a-e.

### Microstructural characterizations

The sample for cross-sectional transmission electron microscopy (TEM) was prepared using a focused ion beam (FIB) system (FEI Helios 5 UC). To minimize ion-beam–induced damage during thinning, Ga$^+$ ions were sequentially applied at accelerating voltages of 30 kV, 5 kV, and 2 kV in the final polishing step. High-resolution structural imaging and elemental analysis were performed using a Thermo Fisher Themis Z probe-corrected TEM operated at an accelerating voltage of 200 kV. High-angle annular dark-field (HAADF) scanning TEM images were acquired using a detector collection angle of 54–200 mrad to enhance Z-contrast between Nb and graphene layers. For energy-dispersive X-ray spectroscopy (EDS) mapping, spectra were recorded with a dwell time of 5 μs per pixel, enabling spatially resolved detection of Nb, C, and O across the junction cross-section.

**Transport measurements**

Transport measurements were performed in a Physical Property Measurement System (PPMS, Quantum Design) equipped with a cryogenic refrigerator capable of reaching 2 K and a superconducting magnet providing magnetic fields up to 9 T. All electrical measurements were carried out using the PPMS Electrical Transport Option (ETO) module.

Current–voltage (IV) characteristics were recorded as a function of magnetic field and temperature to extract the Josephson critical current and normal-state resistance. Differential resistance (conductance) $dV/dI$ ($dI/dV$) maps were obtained by numerically differentiating the IV curves at each field or temperature value. The resulting $dV/dI(I, B)$ and $dI/dV(V, T)$ datasets were used to identify the critical-current envelope, analyze the magnetic-field interference patterns (Fraunhofer patterns), and resolve sub-gap features that might arise from multiple Andreev reflections, respectively.

**References**

[1] Park, J. Y. et al. Double-sided van der Waals epitaxy of topological insulators across an atomically thin membrane. *Nat. Mater.* 24, 399–405 (2025). https://doi.org/10.1038/s41563-024-02079-5

[2] Hong, Y. J. et al. Van der Waals epitaxial double heterostructure: InAs/single-layer graphene/InAs. *Adv. Mater.* 25, 1984–1990 (2013). https://doi.org/10.1002/adma.201302312

[3] Nazir, G. et al. Ultimate limit in size and performance of $WSe_2$ vertical diodes. *Nat. Commun.* 9, 5371 (2018). https://doi.org/10.1038/s41467-018-07820-8

[4] Asshoff, P. U. et al. Magnetoresistance in Co–hBN–NiFe tunnel junctions enhanced by resonant tunneling through single defects in ultrathin hBN barriers. *Nano Lett.* 18, 6954–6960 (2018). https://doi.org/10.1021/acs.nanolett.8b02866

[5] Clarke, John & Wilhelm, Frank K. Superconducting quantum bits. *Nature* 453, 1031–1042 (2008). https://doi.org/10.1038/nature07128

[6] Devoret, M. H. & Schoelkopf, R. J. Superconducting circuits for quantum information: an outlook. *Science* 339, 1169–1174 (2013). https://doi.org/10.1126/science.1231930

[7] Wendin, G. Quantum information processing with superconducting circuits: a review. *Rep. Prog. Phys.* 80, 106001 (2017). https://doi.org/10.1088/1361-6633/aa7e1a

[8] Martinis, J. M., Devoret, M. H. & Clarke, J. Energy-level quantization in the zero-voltage state of a current-biased Josephson junction. *Phys. Rev. Lett.* 55, 1543–1546 (1985). https://doi.org/10.1103/PhysRevLett.55.1543

[9] Oliver, W. D. & Welander, P. B. Materials in superconducting quantum bits. *MRS Bull.* 38, 816–825 (2013). https://doi.org/10.1557/mrs.2013.229

[10] Martinis, J. M. et al. Decoherence in Josephson qubits from dielectric loss. *Phys. Rev. Lett.* 95, 210503 (2005). https://doi.org/10.1103/PhysRevLett.95.210503

[11] Burnett, J. et al. Evidence for interacting two-level systems from the 1/f noise of a superconducting resonator. *Nat. Commun.* 5, 4119 (2014). https://doi.org/10.1038/ncomms5119

[12] Place, A. P. M. et al. New material platform for superconducting transmon qubits with coherence times exceeding 0.3 milliseconds. *Nat. Commun.* 12, 1779 (2021). https://doi.org/10.1038/s41467-021-22030-5

[13] Geim, A. K. & Grigorieva, I. V. Van der Waals heterostructures. *Nature* 499, 419–425 (2013). https://doi.org/10.1038/nature12385

[14] Britnell, L. et al. Field-effect tunneling transistor based on vertical graphene heterostructures. *Science* 335, 947–950 (2012). https://doi.org/10.1126/science.1218461

[15] Lee, G.-H. et al. Ultimately short ballistic vertical graphene Josephson junctions. *Nat. Commun.* 6, 6181 (2015). https://doi.org/10.1038/ncomms7181

[16] Ai, L. et al. Van der Waals ferromagnetic Josephson junctions. *Nat. Commun.* 12, 5487 (2021). https://doi.org/10.1038/s41467-021-26946-w

[17] Kim, M. et al. Strong proximity Josephson coupling in vertically stacked $NbSe_2$–graphene–$NbSe_2$ van der Waals junctions. *Nano Lett.* 17, 6125–6130 (2017). https://doi.org/10.1021/acs.nanolett.7b02707

[18] Balgley, J. et al. Crystalline superconductor-semiconductor Josephson junctions for compact superconducting qubits. *Phys. Rev. Appl.* 24, 034016 (2025). https://doi.org/10.1103/3ssz-jjt6

[19] Wang, J. I-J. et al. Hexagonal boron nitride as a low-loss dielectric for superconducting quantum circuits and qubits. *Nat. Mater.* 21, 398–403 (2022). https://doi.org/10.1038/s41563-021-01187-w

[20] Zhussupbekov, K. et al. Oxidation of Nb(110): atomic structure of the NbO layer and its influence on further oxidation. *Sci. Rep.* 10, 3794 (2020). https://doi.org/10.1038/s41598-020-60508-2

[21] McLellan, R. A. et al. Chemical profiles of the oxides on tantalum in state-of-the-art superconducting circuits. *Adv. Sci.* 10, 2300921 (2023). https://doi.org/10.1002/advs.202300921

[22] Oh, J.-S. et al. Structure and formation mechanisms in tantalum and niobium native oxides in superconducting quantum circuits. *ACS Nano* (2024). https://doi.org/10.1021/acsnano.4c05251

[23] Mizuno, N., Nielsen, B. & Du, X. Ballistic-like supercurrent in suspended graphene Josephson weak links. *Nat Commun.* 4, 2716 (2013). https://doi.org/10.1038/ncomms3716

[24] Chrestin, A. & Merkt, U. High characteristic voltages in Nb/p-type InAs/Nb Josephson junctions. *Appl. Phys. Lett.* 70, 3149–3151 (1997). https://doi.org/10.1063/1.119116

[25] Bessler, R. et al. The dielectric constant of a bilayer graphene interface. *Nanoscale Adv.* 1, 4108–4115 (2019). https://doi.org/10.1039/C8NA00350E

[26] Santos, E. J. G. & Kaxiras, E. Electric-field dependence of the effective dielectric constant in graphene. *Nano Lett.* 13, 898–902 (2013). https://doi.org/10.1021/nl303611v

[27] Kulik, I. O. & Omel'yanchuk, A. N. Contribution to the microscopic theory of the Josephson effect in superconducting bridges. *JETP Lett.* 21, 96 (1975).

[28] Beenakker, C. W. J. Universal limit of critical-current fluctuations in mesoscopic Josephson junctions. *Phys. Rev. Lett.* 67, 3836–3839 (1991). https://doi.org/10.1103/PhysRevLett.67.3836

[29] Likharev, K. K. Superconducting weak links. *Rev. Mod. Phys.* 51, 101–159 (1979). https://doi.org/10.1103/RevModPhys.51.101

[30] Golubov, A. A., Kupriyanov, M. Yu. & Il'ichev, E. The current-phase relation in Josephson junctions. *Rev. Mod. Phys.* 76, 411–469 (2004). https://doi.org/10.1103/RevModPhys.76.411

[31] Dubos, P. et al. Josephson critical current in a long mesoscopic S-N-S junction. *Phys. Rev. B* 63, 064502 (2001). https://doi.org/10.1103/PhysRevB.63.064502

[32] Nikolić, B. K., Freericks, J. K. & Miller, P. Intrinsic reduction of Josephson critical current in short ballistic SNS weak links. *Phys. Rev. B* 64, 212507 (2001). https://doi.org/10.1103/PhysRevB.64.212507

[33] Yang, L. H. et al. Low-energy electron inelastic mean free path and elastic mean free path of graphene. *Appl. Phys. Lett.* 118, 053104 (2021). https://doi.org/10.1063/5.0029133

[34] Staykov, A. & Fujisaki, T. Electron transport through nanoscale multilayer graphene and hexagonal boron nitride junctions. *Beilstein J. Nanotech.* 16, 2132–2143 (2025). https://doi.org/10.3762/bjnano.16.147

[35] Pinto, N. et al. Dimensional crossover and incipient quantum size effects in superconducting niobium nanofilms. *Sci. Rep.* 8, 4710 (2018). https://doi.org/10.1038/s41598-018-22983-6

[36] Afsaneh, E. & Yavari, H. Phase-dependent heat current of granular Josephson junction for different geometries. *Phys. Lett. A* 382, 2388–2393 (2018). https://doi.org/10.1016/j.physleta.2018.05.053

[37] Weihnacht, M. Influence of film thickness on D. C. Josephson current. *Phys. Status Solidi B* 32, 169–174 (1969). https://doi.org/10.1002/pssb.19690320259

[38] Rosenthal, P. A. et al. Flux focusing effects in planar thin-film grain-boundary Josephson junctions. *Appl. Phys. Lett.* 59, 3482–3484 (1991). https://doi.org/10.1063/1.105660

[39] Brandt, E. H. Thin superconductors and SQUIDs in perpendicular magnetic field. *Phys. Rev. B* 72, 024529 (2005). https://doi.org/10.1103/PhysRevB.72.024529

[40] Clem, J. R. Josephson junctions in thin and narrow rectangular superconducting strips. *Phys. Rev. B* 81, 144515 (2010). https://doi.org/10.1103/PhysRevB.81.144515

[41] Hoss, T. et al. Multiple Andreev reflection and giant excess noise in diffusive superconductor/normal-metal/superconductor junctions. *Phys. Rev. B* 62, 4079 (2000). https://doi.org/10.1103/PhysRevB.62.4079

[42] Muñoz, W. A., Covaci, L. & Peeters, F. M. Disordered graphene Josephson junctions. *Phys. Rev. B* 91, 054506 (2015). https://doi.org/10.1103/PhysRevB.91.054506

[43] A. V. Polkin & P. A. Ioselevich, Multiple Andreev reflections in diffusive SINIS and SIFIS junctions. *SciPost Phys.* 15, 100 (2023). https://doi.org/10.21468/SciPostPhys.15.3.100

[44] Zimmermann, U. et al. Multiple Andreev reflection in asymmetric superconducting weak links. *Z. Phys. B* 97, 59–66 (1995). https://doi.org/10.1007/BF01317588

[45] Shu, Z. et al. Near-zero-adhesion-enabled intact wafer-scale resist-transfer printing for high-fidelity nanofabrication on arbitrary substrates. *Int. J. Extrem. Manuf.* 6, 015102 (2024). https://doi.org/10.1088/2631-7990/ad01fe

[46] Song, W. et al. High-Resolution Van der Waals Stencil Lithography for 2D Transistors. *Small* 17, e2101209 (2021). https://doi.org/10.1002/smll.202101209

[47] Schyfter, M. et al. Silicon-barrier Josephson junctions in coplanar and sandwich configurations. *IEEE Trans. Magn.* 13, 862–865 (1977).

[48] Kim, Y. et al. Molecular beam epitaxial step-edge growth of integrated hetero-structures: $Bi_2Te_3$/multi-stepped $Sb_2Te_3$ nanoplate. *2D Mater.* 12, 025008 (2025). https://doi.org/10.1088/2053-1583/ada625

**Acknowledgements**

This work was supported by the National Research Foundation of Korea (NRF) funded by the Korean government (MSIT) (No. RS-2021-NR060087), and by the Science Research Center (SRC) for Novel Epitaxial Quantum Architectures. This research was also supported by the Institute for Basic Science (IBS-R036-D1). Additional support from the National Research Foundation of Korea (Nos. RS-2025-00519093, RS-2021-NR061606 and RS-2024-00356893) is acknowledged. The authors also acknowledge support from the Brain Korea 21 Plus Program, the Institute of Applied Physics (IAP), and the Research Institute of Advanced Materials (RIAM) at Seoul National University.

**Conflict of interest**

The authors declare no competing interests.

**Data availability**

The data that supports the findings of this study are available within the article.

**Author contribution**

Y.K., S.K., J.Y.P., and G.-C.Y. conceived and designed the experiments. Y.K. and S.K. carried out the device fabrication with assistance from S.A., K.J. and J.L. Low-temperature transport measurements were performed by Y.K. and S.K., and the data were analyzed together with J.Y.P. Under the guidance of H.Y., J.K. conducted the TEM-based structural characterization. Y.K. and S.K. wrote the manuscript with input from all co-authors. Y.K. and S.K. contributed equally to this work.

**Figures**

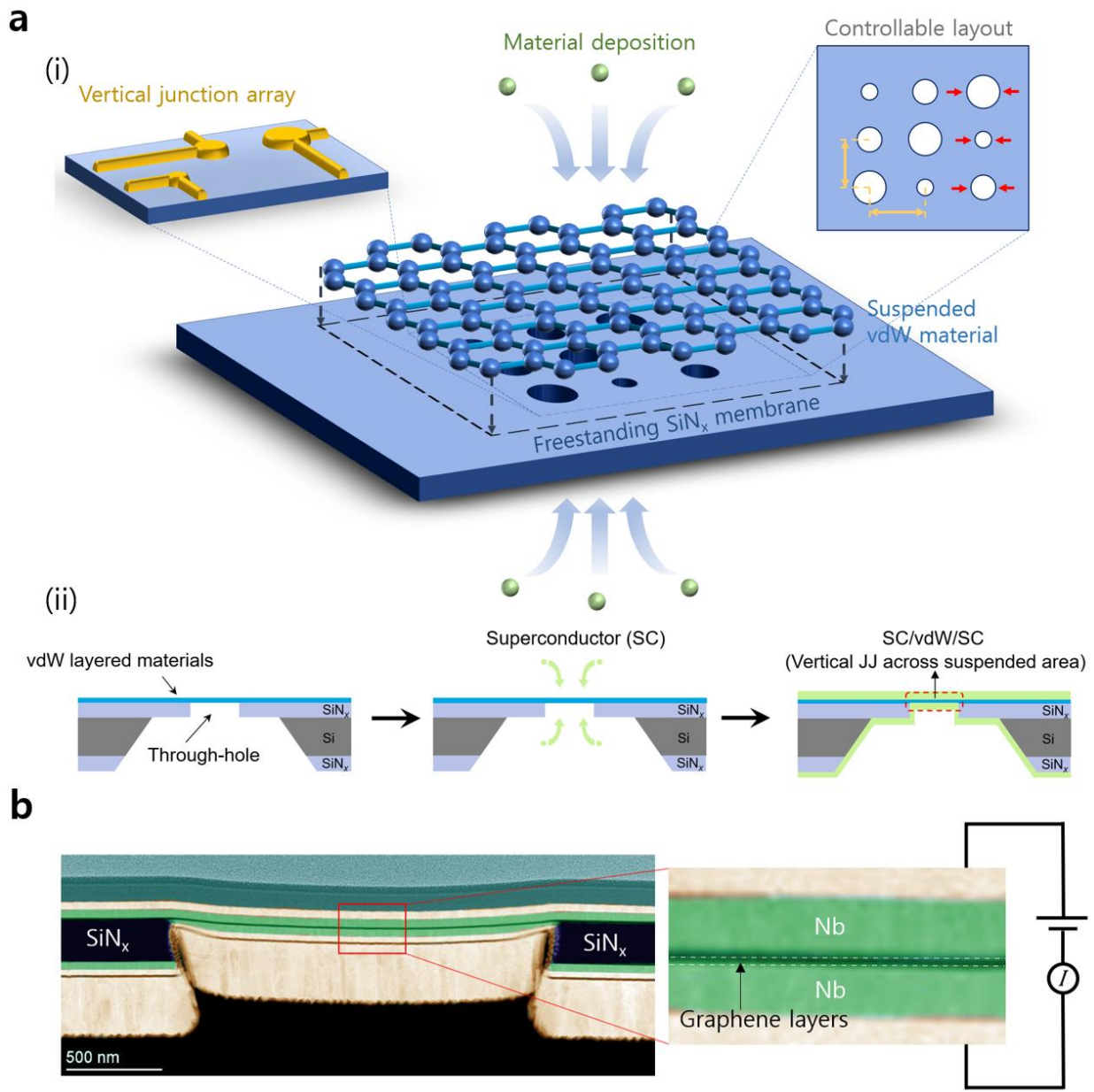

**Figure 1. Concept of freestanding membrane architecture and cross-sectional structure of a vertical Nb/multilayer graphene/Nb Josephson junction.**

**(a)** (i) Schematic of freestanding membrane architecture forming multiple vertical junctions. A suspended van der Waals (vdW) layer spans a lithographically defined through-hole array in a freestanding $SiN_x$ membrane, forming spatially confined active regions accessible from both sides. Since electrical coupling between the top and bottom electrodes occurs exclusively through the suspended region, the active device area is fixed by the aperture geometry, independent of electrode overlap or post-deposition etching. As a result, processing on the top and bottom surfaces of the membrane can be carried out independently: bottom electrodes may be deposited over large areas, while the number, spacing, and density of vertical junctions are solely defined by the aperture layout and pitch on the top side. (ii) Cross-sectional schematic of the freestanding membrane approach used in this work. As illustrated, a vdW layered material is first suspended over a lithographically defined through-hole array in a $SiN_x$ membrane. Superconducting (SC) layers are then sequentially deposited on both sides of the suspended region, forming vertical SC/vdW/SC Josephson junctions across the membrane apertures. In contrast, in conventional bottom-up fabrication schemes, a pre-deposited SC layer would be degraded during subsequent processing required to suspend a vdW layer above it. This degradation prevents the realization of a vertical Josephson junction, even if an additional SC layer is later deposited on top of the vdW material. **(b)** False-coloured low magnification cross-sectional bright-field TEM image of a vertical junction. A multilayer graphene sheet spans the $SiN_x$ membrane opening and is sandwiched between Nb layers deposited on the top and bottom surfaces. The magnified inset highlights the graphene layers located between the two Nb electrodes. A schematic symbol on the right indicates the electrical measurement configuration.

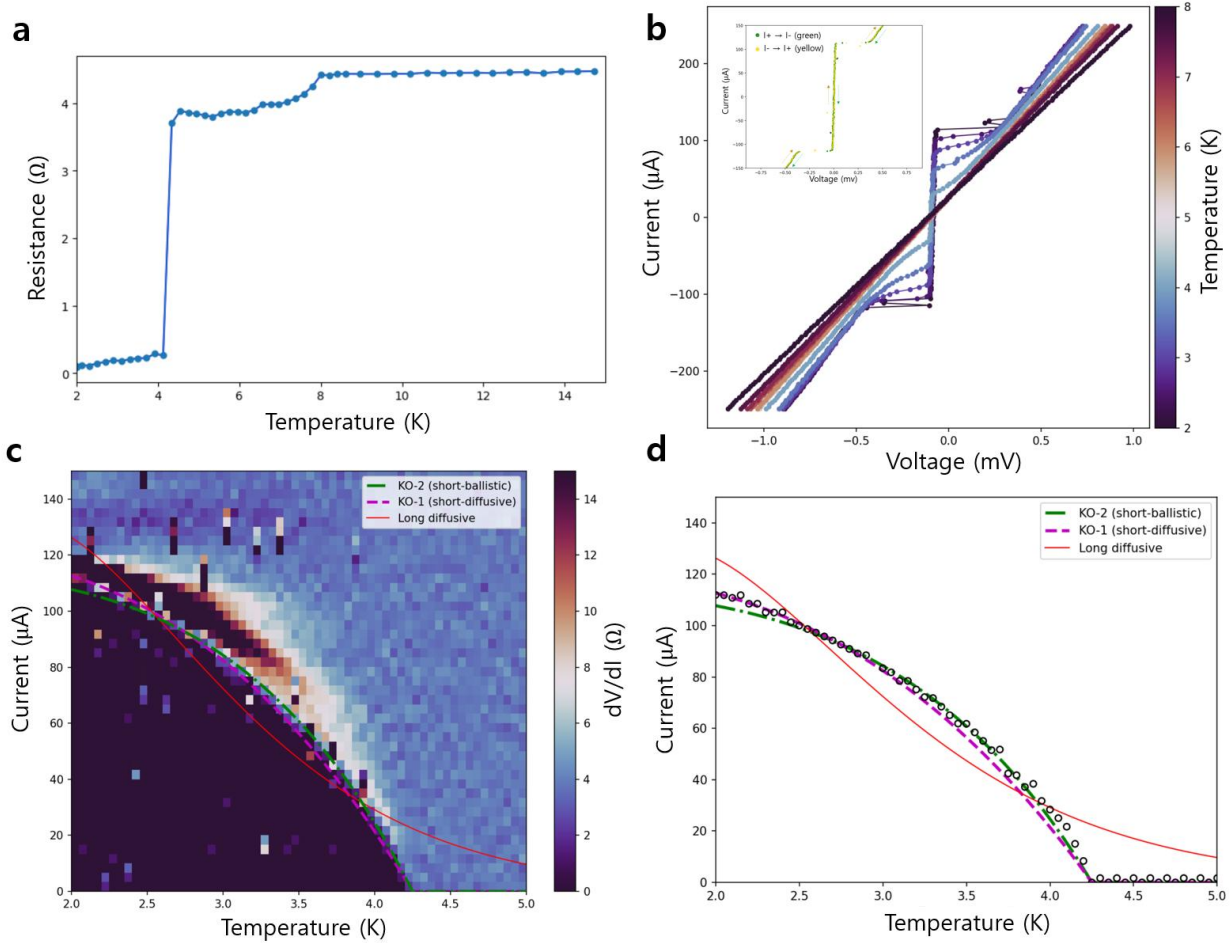

**Figure 2. Josephson coupling and temperature-dependent transport in a vertical Nb/multilayer graphene/Nb junction.**

(**a**) Temperature dependence of the junction resistance. The resistance gradually decreases below ~8 K, reflecting the superconducting transitions of the top and bottom Nb electrodes, and then drops sharply to nearly zero below ~4.3 K, marking the onset of Josephson coupling. (**b**) Current–voltage (IV) characteristics measured at temperatures between 2 K and 8 K. The supercurrent decreases monotonically with increasing temperature, from ~110 μA at 2 K to zero above ~4.3 K ($T_c$ of the Josephson junction). IV characteristics measured at 2 K for opposite current sweep directions are shown as an inset (upper left). The curves exhibit negligible hysteresis in the supercurrent region, consistent with the SNS-type Josephson behaviour expected for the vertical Nb/multilayer graphene/Nb junction. (**c**) Two-dimensional color map of differential resistance (*dV/dI*) as a function of bias current and temperature. The boundary between the zero-resistance and resistive regions defines the critical current. The overlaid curves correspond to fits using three standard models—short-ballistic (KO-2), short-diffusive (KO-1), and long-diffusive limits—shown in green dashed, purple dashed, and red solid lines, respectively. **(d)** Extracted critical current as a function of temperature (symbols), together with theoretical fits based on the short-ballistic (KO-2), short-diffusive (KO-1), and long-diffusive models.

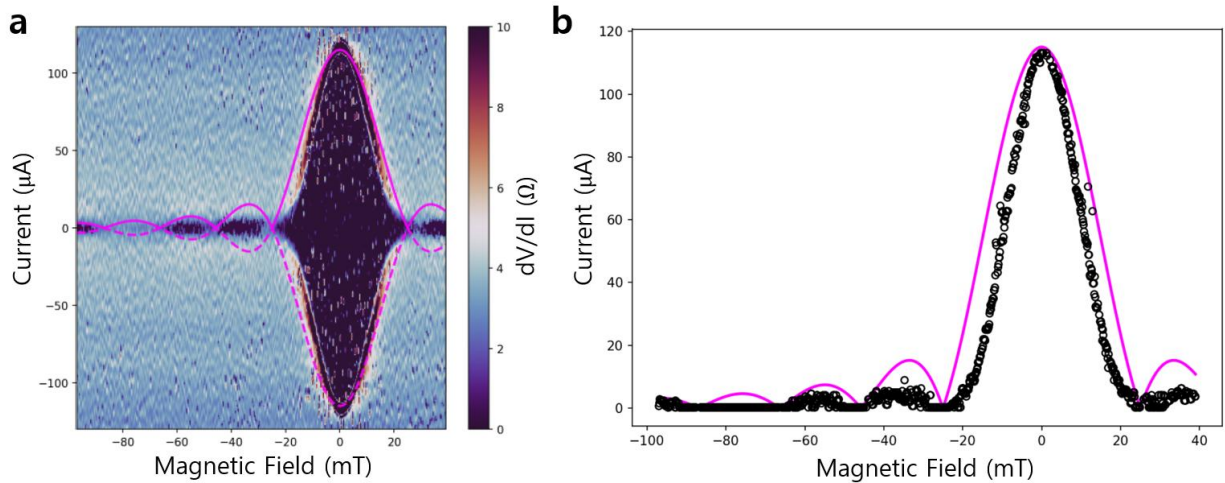

**Figure 3. Magnetic-field interference (Fraunhofer) pattern of the vertical Josephson junction.**

**(a)** Differential resistance *dV/dI* plotted as a function of bias current and magnetic field applied parallel to the junction plane. A clear Fraunhofer interference pattern with a periodicity of ~25 mT is observed. The overlaid magenta curves represent the expected envelope for a circular (cylindrical) junction geometry defined by the membrane through-hole. **(b)** Critical current extracted from the *dV/dI* map as a function of magnetic field. The experimental data are overlaid with the same circular-aperture Fraunhofer envelope, showing good agreement with the cylindrical junction geometry.

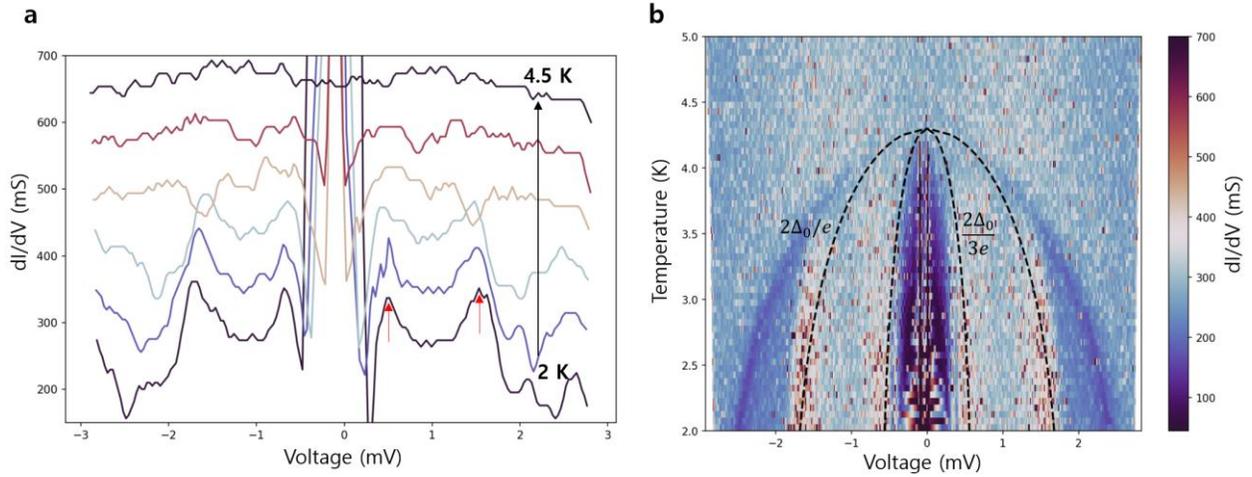

**Figure 4. Sub-gap features in the Nb/multilayer graphene/Nb vertical Josephson junction.**

(**a**) Differential conductance *dI/dV* as a function of voltage, measured at several temperatures between 2 K and 4.5 K. At 2 K, pronounced sub-gap structures are visible, and the positions of the features suggestive of multiple Andreev reflections (MAR) are marked by red arrows, corresponding to the $n = 1$ and $n = 3$ MAR orders. (**b**) Two-dimensional map of *dI/dV* plotted as a function of voltage and temperature. The dashed curves indicate the expected MAR peak positions obtained from BCS theory, showing the voltage scales $V_1 = 2\Delta_0/e$ and $V_3 = 2\Delta_0/3e$ associated with the $n = 1$ and $n = 3$ MAR processes. The evolution of these sub-gap features with temperature is consistent with the suppression of the superconducting gap.

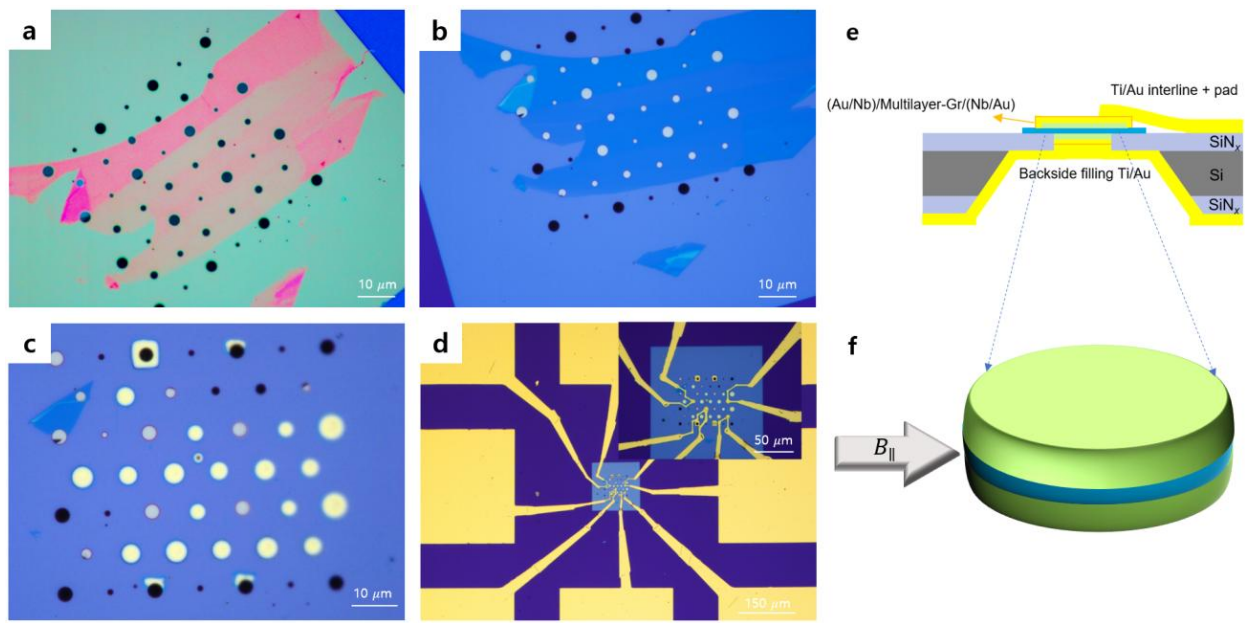

**Extended Data Figure 1. Josephson junction fabrication processes and schematic diagrams.**

**(a)-(d)** Optical microscope images illustrating the sequential fabrication processes of Josephson junctions. **(a)** Multilayer graphene is first transferred onto a freestanding $SiN_x$ membrane patterned with a through-hole array, followed by chemical and thermal cleaning steps. **(b)** Backside deposition of Nb/Au (47/20 nm) by sputtering, followed by Ti/Au (10/240 nm) deposition via e-beam evaporation, results in the filling of the through-holes beneath the regions covered by multilayer graphene. **(c)** Optical image of the device after top-side Nb/Au (47/30 nm) deposition by sputtering and subsequent reactive ion etching (RIE) to define individual top contacts. **(d)** Optical image of the final device, including interconnecting lines and bonding pads formed by Ti/Au (10/140 nm). **(e), (f)** Cross-sectional schematic of the final device geometry and a schematic illustration of the externally applied magnetic field oriented parallel to the in-plane direction of the multilayer graphene.

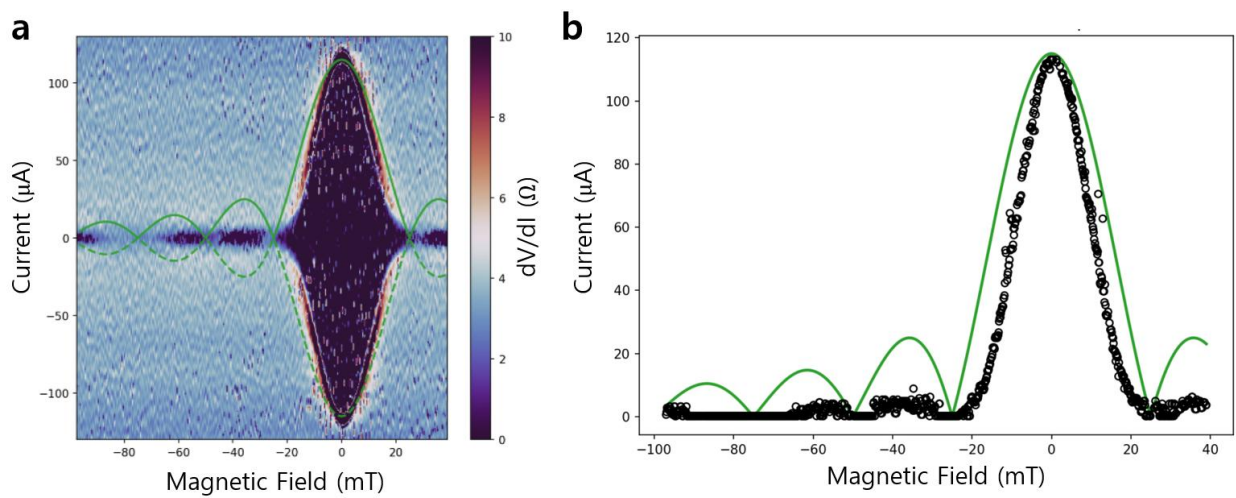

**Extended Data Figure 2. Comparison of Fraunhofer patterns with a rectangular junction geometry model.**

**(a)** Experimental *dV/dI* map as a function of bias current and magnetic field, showing a well-defined Fraunhofer interference pattern with a periodicity of ~25 mT, overlaid with the envelope (green) expected for a rectangular junction of equal nominal area. **(b)** Critical current extracted from the map in **(a)**, compared with the same rectangular-junction envelope. The rectangular model fails to reproduce the experimentally observed node positions, in contrast to the circular-aperture model shown in Fig. 3.

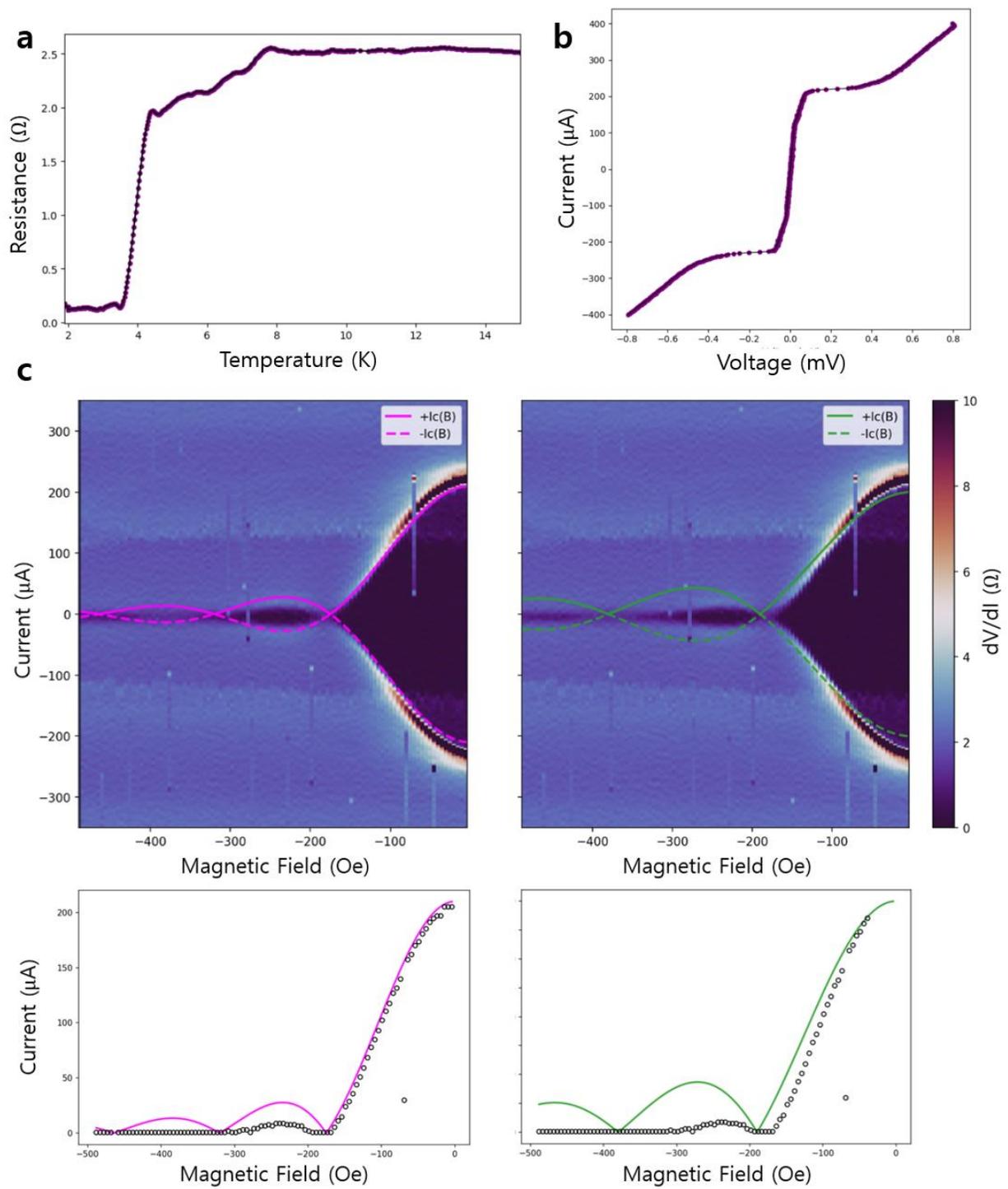

**Extended Data Figure 3. Reproducibility of Josephson characteristics in an additional device.**

**(a)** Temperature-dependent resistance of an additional Nb/multilayer graphene/Nb junction fabricated on the membrane platform. The superconducting transition and overall resistance evolution closely resemble those of the main device. **(b)** Current–voltage characteristics measured at 2 K, showing a clear supercurrent branch. **(c)** Magnetic-field-dependent differential-resistance maps with a periodicity of ~17.5 mT, consistent with the larger effective junction area of this device. The left column shows the experimental *dV/dI* map and the extracted critical current as a function of magnetic field, overlaid with the envelope (magenta) expected for a cylindrical junction geometry defined by the through-hole aperture. The right column shows a comparison with the rectangular-junction envelope (green), which fails to capture the observed node positions. These measurements demonstrate that the key Josephson characteristics are reproducible across multiple devices.

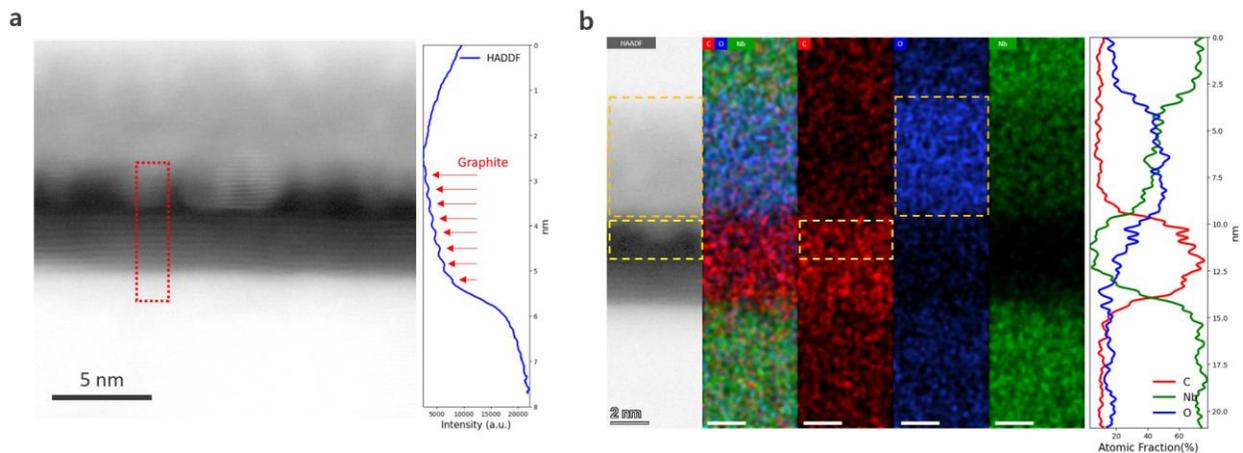

**Extended Data Figure 4. Structural and compositional analysis of a non-functional junction.**

**(a)** High-angle annular dark-field scanning transmission electron microscopy (HAADF-STEM) image of the Nb/8L-graphene/Nb stack. The layered contrast of the transferred 8-layer graphene is visible in the highlighted region, while additional contrast near the top Nb interface suggests the presence of an unintended interfacial layer. The line-intensity profile on the right clearly resolves the individual graphene layers. **(b)** Elemental maps of the same cross-section acquired by energy-dispersive X-ray spectroscopy (EDS). The yellow boxes denote regions where carbon-rich signals extend into the upper region above the graphene, outside the intrinsic graphene stack, while the orange boxes highlight areas where oxygen and Nb signals coexist near the top interface. The atomic-fraction profiles on the right confirm localized enrichment of C above the graphene region and identify a region of overlapping O and Nb signals, consistent with the formation of a C-O-Nb interfacial layer. In contrast, the graphene/bottom-Nb interface exhibits neither an oxygen-rich region nor anomalous carbon signals.